\input{aipcheck}

\documentclass[
    ,final            
  ]
  {aipproc}

\layoutstyle{6x9}

\begin{document}

\title{Some Hadronic Properties \\ from Light Front Holography}


\author{Alfredo Vega}{
  address={Departamento de F\'\i sica y Centro Cient\'\i
fico Tecnol\'ogico de Valpara\'\i so (CCTVal), Universidad T\'ecnica
Federico Santa Mar\'\i a, Casilla 110-V, Valpara\'\i so, Chile}
}

\author{Ivan Schmidt}{
  address={Departamento de F\'\i sica y Centro Cient\'\i
fico Tecnol\'ogico de Valpara\'\i so (CCTVal), Universidad T\'ecnica
Federico Santa Mar\'\i a, Casilla 110-V, Valpara\'\i so, Chile}
}

\author{Thomas Gutsche}{
  address={Institut f\"ur Theoretische Physik,
Universit\"at T\"ubingen, \\
Kepler Center for Astro and Particle Physics,
\\ Auf der Morgenstelle 14, D-72076 T\"ubingen, Germany} 
}

\author{\\ Valery E. Lyubovitskij 
\thanks{
On leave of absence
from Department of Physics, Tomsk State University,
634050 Tomsk, Russia}
}{
  address={Institut f\"ur Theoretische Physik,
Universit\"at T\"ubingen, \\
Kepler Center for Astro and Particle Physics,
\\ Auf der Morgenstelle 14, D-72076 T\"ubingen, Germany}
}

\begin{abstract}
 Using ideas from Light Front Holography, we discuss the calculation of hadronic properties. In this talk I will pay special attention to hadronic masses and the nucleon helicity-independent generalized parton distributions of quarks in the zero skewness case.
\end{abstract}

\maketitle


\section{Introduction}

~~~~Light Front Holography (LFH)~\cite{LFH,Brodsky:2008pg} is a semiclassical approximation to QCD based on the AdS / CFT correspondence. It provides a precise mapping of the string modes $\Phi(z)$ in the AdS fifth dimension $z$ to the hadron light-front wave functions (LFWFs) in physical space-time, and this approach has been successfully applied to the description of the mass spectrum of mesons and baryons (reproducing the Regge trajectories), the pion leptonic constant, the electromagnetic form factors of pion and nucleons, etc.~\cite{LFH,Brodsky:2008pg,Vega:2009zb, Branz:2010ub}. 

The mapping that allows to relate AdS modes with LFWF was obtained by matching certain matrix elements (e.g. electromagnetic pion form factor) in two approaches - string theory in AdS and Light-Front QCD in Minkowski space-time, and the same ideas can be used to calculate generalized parton distributions (GPDs)~\cite{Vega:2010ns}.


\section{Mesonic Phenomenology}

~~~~Comparing mesonic form factors calculated in light front QCD and using AdS / CFT  ideas, it is possible to obtain a relation between mesonic wave function (WF) and AdS modes (for details see ~\cite{Vega:2009zb, Branz:2010ub}): 
\begin{equation}
\label{LFWF}
| \psi (x,\zeta) |^{2} = A \frac{1}{\zeta} x (1-x) f(x) |\Phi(\zeta)|^{2}.
\end{equation}
where A is a normalization constant, $x$ is the light front momentum fraction, $z=\zeta$ and $\Phi(\zeta)$ 
is the solution of the Schr\"odinger equation 
\begin{equation}
\Big[ - \frac{d^2}{dz^2} + U_J(z) \Big] \Phi_{nJ}(z) = M^2_{nJ} \Phi_{nJ}(z)
\end{equation}
in $z$-dimension of AdS$_{d+1}$ space-time with
effective potential $U_J(z)$ depending on quadratic dilaton field. 
Here variable $\zeta$ is related to impact variable ${\bf b}_\perp$ 
as $\zeta=\sqrt{{\bf b}_\perp^2 x (1-x)}$. 
Extracting mesonic WF in impact space from (\ref{LFWF}) and performing 
a Fourier transformation we derive the mesonic  WF in momentum space: 
\begin{equation}
\psi_{q_1 \bar{q}_2}(x,k) = \frac{4\pi A}{\kappa \sqrt{x(1-x)}} 
\exp\biggl(-\frac{k^2}{2 \kappa^2 x(1-x)}\biggr).
\end{equation}
It is possible to extend this model in order to include massive quarks 
($m_{1}$ and $m_{2}$), extending the kinetic 
energy of the two massless constituents by 
a mass term~\cite{Brodsky:2008pg,Vega:2009zb,Branz:2010ub}: 
\begin{equation}
\frac{k^{2}}{x(1-x)} \to \frac{k^{2}}{x(1-x)} + m_{12}^2~~~,~~~m_{12}^2 = \frac{m^2_1}{x} + \frac{m^2_2}{1-x}.
\end{equation}
Therefore, the meson WF with massive quarks in momentum space is
\begin{equation}
\psi_{q_1 \bar{q}_2}(x,k) = \frac{4\pi}{\kappa \sqrt{x(1-x)}} f(x,m_{1},m_{2}) \exp \biggl(-\frac{k^2}{2 \kappa^2 x(1-x)}\biggr).
\end{equation} 
with 
\begin{equation} 
f(x,m_{1},m_{2})=A~f(x) e^{-\frac{m_{12}^2}{2 \lambda^{2}}}
\end{equation}
An additional extension we included extra terms in the effective potential, 
i.e $U \to U + U_{\rm C} + U_{\rm HF}$, 
where $U_{\rm C}$ and $ U_{\rm HF}$ are the contributions of 
the color Coulomb and hyperfine (HF) potentials. 
See further details and applications in~\cite{Branz:2010ub}. 
Here, in Table 1 we just show results for masses 
of heavy-light mesons. 

\begin{table}
\caption{Masses of heavy-light mesons.}
\begin{tabular}{c|c|c|c|c|c|c|c|c}
\hline
Meson&$J^{\rm P}$&$n$&$L$&$S$&\multicolumn{4}{c}{Mass [MeV]} \\
\hline
$D(1870)$&$0^{-}$&0&0,1,2,3         &0& 1857 & 2435 & 2696 & 2905 \\ \hline
$D^{\ast}(2010)$&$1^{-}$&0&0,1,2,3  &1& 2015 & 2547 & 2797 & 3000 \\ \hline
$D_s(1969)$&$0^{-}$&0&0,1,2,3       &0& 1963 & 2621 & 2883 & 3085 \\ \hline
$D^{\ast}_s(2107)$&$1^{-}$&0&0,1,2,3&1& 2113 & 2725 & 2977 & 3173 \\ \hline
$B(5279)$&$0^{-}$&0&0,1,2,3         &0& 5279 & 5791 & 5964 & 6089 \\ \hline
$B^{\ast}(5325)$&$1^{-}$&0&0,1,2,3  &1& 5336 & 5843 & 6015 & 6139 \\ \hline
$B_s(5366)$&$0^{-}$&0&0,1,2,3       &0& 5360 & 5941 & 6124 & 6250 \\ \hline
$B^{\ast}_s(5413)$&$1^{-}$&0&0,1,2,3&1& 5416 & 5992 & 6173 & 6298 \\ \hline
\end{tabular}
\end{table}


\section{Generalized Parton Distributions}

~~~~We consider a matching of the nucleon electromagnetic form factors in two approaches \cite{Vega:2010ns}: we use sum rules derived in QCD~\cite{GPD}, which contain GPDs for valence quarks, and we also consider an expression obtained in the AdS/QCD soft-wall model~\cite{Abidin:2009hr}. As result of the matching we obtain the expressions for the nonforward parton densities $H_{v}^{q}(x,t) = H^q(x,0,t) + H^q(-x,0,t)$ and $E_{v}^{q}(x,t) = E^q(x,0,t) + E^q(-x,0,t)$ -- flavor combinations of the GPDs (or valence GPDs), 
using information obtained on the AdS side:  
\begin{equation}
H_{v}^{q}(x,Q^2) = q(x) \, x^a \,, \quad 
E_{v}^{q}(x,Q^2) = e^q(x) \, x^a \,. 
\end{equation}
Here $q(x)$ and $e^q(x)$ are distribution functions given by: 
\begin{equation}
q(x)   = \alpha^q \gamma_{1}(x) + \beta^q \gamma_{2}(x)\,, \quad 
e^q(x) = \beta^q \gamma_{3}(x)\,, 
\end{equation}
where the flavor couplings $\alpha_q, \beta_q$ and functions $\gamma_i(x)$ are written as
\begin{equation}
\alpha^u = 2\,, \ \alpha^d = 1\,, \ 
\beta^u = 2 \eta_{p} + \eta_{n} \,, \ 
\beta^d = \eta_{p} + 2 \eta_{n} \,  
\end{equation}
and 
\begin{equation}
\gamma_{1}(x) =
\frac{1}{2} (5 - 8x + 3x^{2})\,, \nonumber\\
\gamma_{2}(x) = 1 - 10x + 21x^{2} - 12x^{3} \,, 
\label{gamma} \\
\gamma_{3}(x) = 
\frac{6 m_N \sqrt{2}}{\kappa} (1 - x)^{2} \,. \nonumber
\end{equation}
Expressions for the GPDs in terms of the AdS modes can be obtained using  
the LFH procedure of mapping, and are plotted in Fig 1 \cite{Vega:2010ns}. 
\begin{figure}[ht]
  \begin{tabular}{c c}
    \includegraphics[width=3.0 in]{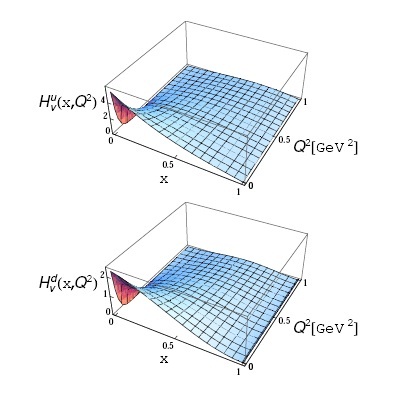} & \includegraphics[width=3.0 in]{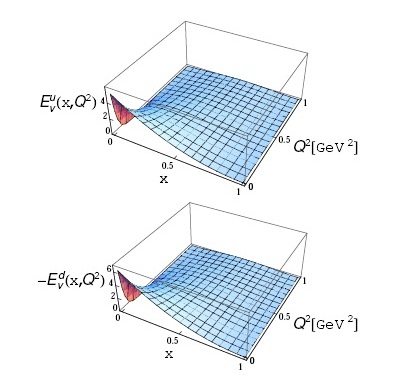}
  \end{tabular}
\caption{GPDs $H_{v}^{q} (x,Q^2)$ and $E_{v}^{q} (x,Q^2)$ calculated in the holographical model.}
\end{figure}


\section{Conclusions}

~~~~The AdS/CFT correspondence between Anti-de Sitter space and conformal gauge theory provides an analytically tractable approximation to QCD called light front holography. Matching certain matrix elements (e.g. electromagnetic pion form factor) in the two approaches - string theory in AdS and Light-Front QCD in Minkowski space-time, allow us to obtain light-front wavefunctions of mesons and GPDs of baryons, the fundamental entities which encode hadronic properties and allow the computation of some hadronic properties.


\begin{theacknowledgments}
~~~~The authors thanks Stan Brodsky, Guy de T\'eramond and Tanja Branz for useful discussions. This work was supported by Federal Targeted Program "Scientific and scientific-pedagogical personnel of innovative Russia" Contract No. 02.740.11.0238, by FONDECYT (Chile) under Grant No. 1100287. A. V. acknowledges the financial support from FONDECYT (Chile) Grant No. 3100028.
\end{theacknowledgments}



\bibliographystyle{aipproc}   


\IfFileExists{\jobname.bbl}{}
 {\typeout{}
  \typeout{******************************************}
  \typeout{** Please run "bibtex \jobname" to optain}
  \typeout{** the bibliography and then re-run LaTeX}
  \typeout{** twice to fix the references!}
  \typeout{******************************************}
  \typeout{}
 }


\end{document}